\documentclass[aps,pra,preprint]{revtex4-1}%
\usepackage{amsfonts}
\usepackage{amsmath}
\usepackage{amssymb}
\usepackage{graphicx}

\begin{document}
\title{Quantum capacitance of an ultrathin topological insulator film in a
magnetic field}
\author{M. Tahir$^{1}$, K. Sabeeh$^{2}$, and U. Schwingenschl\"{o}gl$^{1,}$}
\email{Correspondence to udo.schwingenschlogl@kaust.edu.sa}
\affiliation{$^{1}$PSE Division, KAUST, Thuwal 23955-6900, Kingdom of Saudi Arabia}
\affiliation{$^{2}$Department of Physics, Quaid-i-Azam University, Islamabad 45320 Pakistan}

\begin{abstract}
We present a theoretical study of the quantum magnetocapacitance of an ultrathin
topological insulator film in an external magnetic field. The
study is undertaken to investigate the interplay of the Zeeman interaction with
the hybridization between the upper and lower surfaces of the thin film.
Determining the density of states, we find that the electron-hole symmetry is
broken when the Zeeman and hybridization energies are varied relative to each other. 
This leads to a change in the character of the magnetocapacitance at the charge
neutrality point. We further show that in the presence of both Zeeman interaction
and hybridization the magnetocapacitance exhibits beating at low
and splitting of the Shubnikov de Haas oscillations at high perpendicular magnetic field.
In addition, we address the crossover from perpendicular to parallel magnetic field
and find consistency with recent experimental data.
\end{abstract}

\maketitle


Topological insulators (TIs) form a new class of compounds with strong
spin-orbit interaction \cite{1,2,3,4,5,6,7,8}. These materials possess an energy gap
in the bulk (like insulators) and conducting electronic states at the surface (like
metals). The nature of the surface states depends on the dimensions of the TI
\cite{9,new4,new5,new6,11,12,13,14}: Two-dimensional surface states have been revealed in
transport experiments for three-dimensional TIs. The charge carriers in such
surface states form a two-dimensional gas of Dirac fermions with a single
Dirac cone energy spectrum. Dirac fermions have been observed
in angle resolved photo-emission experiments and quantization of their Landau
levels (LLs) has been confirmed by scanning tunneling spectroscopy \cite{stm1,stm2,stm3,stm4,stm5}.
TI thin films provide a new degree of freedom which is the thickness of the film. Effects of the
crossover from 3D to 2D topological surface states as well as hybridization of
the states at the two surfaces of the thin film have been studied for varying
film thicknesses and found to become important when the thickness is less than about
5 nm \cite{16,17,18,19,20}. The Dirac fermions in TIs share several
features with those of graphene \cite{21,22}. However, there are important
differences between graphene and the surface states of a TI. Unlike graphene,
there is an odd number of not spin degenerate Dirac points at any surface.
In addition, TIs exhibit a unique spin-momentum locking in the surface states.

Observation of magnetic quantum oscillations, Shubnikov de Haas (SdH) and de
Haas van Alphen \cite{23,24}, in the transport experiments has resulted
in theoretical work on magnetization \cite{23} and Berry
phase \cite{24,25} features of TI surface Dirac fermions. Furthermore, much
attention is paid to magnetically doped TIs due to their exceptional
properties and potential spintronics applications \cite{12,18}. In this
context, capacitance measurements are among the most important tools for
studying the electronic properties. They can be effectively used to probe the thermodynamic
density of states (DOS). Although the focus in TI research
has been on transport properties, insight into the fundamental electronic
properties and device physics still calls for knowledge about the
capacitance-voltage (C-V) characteristics of the system. Results have been reported
for carbon nanotubes, graphene nanoribbons, and mono- and bilayer
graphene systems \cite{26,27,28,29,30,31}. In view of this, attention is now
being paid to electrostatic properties, such as the magnetocapacitance of TI thin
film devices. Furthermore, for improve the
performance of field effect transistors \cite{32}, the potential of TI
thin films as channel materials is creating much excitement. This
is due to the excellent intrinsic transport features as well as the
possibility of patterning device structures within top-down lithographical
techniques. The present work aims at determining the combined effects of the Zeeman
interaction and hybridization on the magnetocapacitance of a TI thin film device.

\section{Results}

We consider Dirac fermions on the surface of a 3D TI thin film which we take
to be in the $xy$-plane. There is an external magnetic field perpendicular to
the thin film and we take into account hybridization between the upper and
lower surfaces. The two-dimensional Hamiltonian for Dirac fermions
in a magnetic field with hybridization is given by \cite{18,21,22}
\begin{equation}
H=v(\sigma_{x}\boldsymbol{\pi}_{y}-\tau_{z}\sigma_{y}\boldsymbol{\pi}_{x})+(\tau_{z}\Delta
_{z}+\Delta_{h})\sigma_{z}. \label{1}%
\end{equation}
Here, $\sigma_{x}$ and $\sigma_{y}$ are
Pauli matrices that operate on a combined space of spin and surface degrees of freedom,
$\tau_{z}=+/-$ denotes symmetric/antisymmetric combinations of the upper and lower surfaces
of the TI, $v$ denotes the Fermi velocity of the
Dirac fermions, $\boldsymbol{\pi}=\mathbf{p}+e\mathbf{A}/c$ is the
two-dimensional canonical momentum with vector potential $\mathbf{A}$, and $c$
is the speed of light. Moreover, we have the Zeeman energy $\Delta_{z}=\frac
{1}{2}g\mu_{B}B$, the effective Lande factor $g$, the Bohr magneton $\mu_{B}$,
and the hybridization matrix element $\Delta_h$, which reflects the
hybridization between the upper and lower surfaces of the TI. We employ
the Landau gauge and express the vector potential as $\mathbf{A}=(0,Bx,0)$.
The LL energies then are given by
\begin{align}
E_{0}^{\tau_{z}}  &  =-(\Delta_{z}+\tau_{z}\Delta_{h})\label{2}\\
E_{n,\lambda}^{\tau_{z}}  &  =\lambda\sqrt{2n\hslash^{2}\omega^{2}+(\Delta
_{z}+\tau_{z}\Delta_{h})^{2}},\quad n\neq0\nonumber
\end{align}
with $\lambda=+/-$ for electron/hole bands. Moreover, $\omega=v\sqrt{eB/\hslash}$ is
the cyclotron frequency of the Dirac fermions and $n$ is an integer. The main
point to note about the energy spectrum, which will be crucial to the following
discussion, is: The splitting of the energy levels depends on
both the Zeeman interaction and the hybridization between the upper and lower
surfaces. If either of these two mechanisms is absent, no splitting
will occur. Further, the energy spectrum is electron-hole symmetric in
the absence of the Zeeman interaction, with a hybridization gap.
We note that angular resolved photoelectron spectroscopy finds no electron-hole
symmetry even in the absence of both Zeeman interaction and hybridization, reflecting
a quadratic term in the Hamiltonian \cite{rev1,rev2}. However, in our surface
Hamiltonian in Eq.\ (2) no quadratic term is present as it can be neglected when
the system is doped such that the Dirac point comes close to the charge neutrality
point, which is the focus of our work. In the presence of both Zeeman interaction
and hybridization, the $n=0$ LL explicitly shows a quantum phase transition. This
will be important in the discussion of the DOS and the capacitance results.

We consider a gated TI device in which the capacitor is formed between the
gate and the TI thin film. The magnetocapacitance of the device represents
the charge response in the channel as the channel potential is varied. In
conventional devices, the magnetocapacitance $C_{Q}$ is large and can be
ignored. On the contrary, in low dimensional devices, such as 2D electron gases,
graphene, and future TI devices, it is the dominant capacitive contribution and therefore
an important quantity in the design of nanoscale devices. The central
expression for the quantum capacitance is \cite{26}
\begin{equation}
C_{Q}=\frac{e\partial Q}{\partial\varepsilon_{F}}=\frac{e^{2}\partial n_{e}%
}{\partial\varepsilon_{F}}=e^{2}D_{T}(B), \label{4}%
\end{equation}
where $n_{e}$ is the carrier concentration and $\varepsilon_F$ the Fermi energy.
The temperature dependent DOS at a
finite magnetic field, $D_{T}(B)$, is obtained from the relation%
\begin{equation}
D_{T}(B)=\frac{\partial n_{e}}{\partial\varepsilon_{F}}=
{\displaystyle\int\limits_{0}^{\infty}}
d\varepsilon\frac{\partial f(\varepsilon-\varepsilon_{F})}{\partial
\varepsilon_{F}}D(\varepsilon), \label{5}%
\end{equation}
with the Fermi Dirac distribution $f$. In the limit of zero temperature we have
$D_{T}(B)=D(\varepsilon_{F})$ and
\begin{equation}
D(\varepsilon_{F})=\frac{1}{2\pi l^{2}}\underset{n,\tau_{z},\lambda}{%
{\displaystyle\sum}
}\delta\left(  \varepsilon_{F}-E_{n,\lambda}^{\tau_{z}}\right)  . \label{6}%
\end{equation}
Assuming a Gaussian broadening of the LLs, the DOS per unit area is given by
\begin{equation}
D(\varepsilon_{F})=\frac{1}{2\pi l^{2}}\underset{n,\tau_{z},\lambda}{%
{\displaystyle\sum}
}\frac{1}{\Gamma\sqrt{2\pi}}\exp\left[  -\frac{(\varepsilon_{F}-E_{n,\lambda
}^{\tau_{z}})^{2}}{2\Gamma^{2}}\right], \label{7}%
\end{equation}
where $\Gamma$\ is the width of the Gaussian distribution (zero shift).
For $n=0$ we have
\begin{equation}
D_{n=0}(\varepsilon_{F})=\frac{1}{2\pi l^{2}}\underset{\tau
_{z}}{%
{\displaystyle\sum}
}\frac{1}{\Gamma\sqrt{2\pi}}\exp\left[  -\frac{(\varepsilon_{F}-E_{0}%
^{\tau_{z}})^{2}}{2\Gamma^{2}}\right], \label{8}%
\end{equation}
where $E_{0}^{s}=-(\Delta_{z}+\tau_{z}\Delta_{h})$. Combining Eqs.\ (8) and (9), we obtain
\begin{equation}
D(\varepsilon_{F})=\frac{1}{2\pi l^{2}}\left\{  \underset{\tau_{z}}{%
{\displaystyle\sum}
}\frac{1}{\Gamma\sqrt{2\pi}}\exp\left[  -\frac{(\varepsilon_{F}-E_{0}%
^{\tau_{z}})^{2}}{2\Gamma^{2}}\right]  +\underset{\tau_{z},\lambda
,n=1}{\overset{\infty}{%
{\displaystyle\sum}
}}\frac{1}{\Gamma\sqrt{2\pi}}\exp\left[  -\frac{(\varepsilon_{F}-E_{n,\lambda
}^{\tau_{z}})^{2}}{2\Gamma^{2}}\right]\right\}.  \label{9}%
\end{equation}
At the charge neutrality point ($\varepsilon_{F}=0$) the result is
\begin{equation}
D(\varepsilon_{F}=0)=\left(  \frac{1}{2\pi l^{2}}\underset{\tau_{z}}{%
{\displaystyle\sum}
}\frac{1}{\Gamma\sqrt{2\pi}}\exp\left[  -\frac{(\Delta_{z}+\tau_{z}\Delta
_{h})^{2}}{2\Gamma^{2}}\right]  \left\{  1+2\underset{n=1}{\overset{\infty}{%
{\displaystyle\sum}
}}\exp\left[  -\frac{2n\hslash^{2}\omega^{2}}{2\Gamma^{2}}\right]  \right\}
\right), \label{10}%
\end{equation}
which can be written as
\begin{equation}
D(\varepsilon_{F}=0)=\left(  \frac{\Gamma}{\left(  V_{F}\hslash\right)
^{2}\pi\sqrt{2\pi}}\underset{\tau_{z}}{%
{\displaystyle\sum}
}\exp\left[  -\frac{(\Delta_{z}+\tau_{z}\Delta_{h})^{2}}{2\Gamma^{2}}\right]
\left\{  \frac{\chi}{\tanh\chi}\right\}  \right)  \label{11}%
\end{equation}
where $\chi=\hslash^{2}\omega^{2}/2\Gamma^{2}$.
Equation (10) reduces to the result reported in Ref.\ \cite{26} in the
absence of Zeeman interaction and hybridization.

The magnetocapacitance $C_{Q}$ is plotted in Figs.\ 1 and 2 as a function of
the Fermi energy (i.e., of the gate voltage). The following parameters are employed
\cite{new1,new2,new3,16,24}:
$g=60$, $B=3$ T, $\Delta_{z}=5$ meV, and $\Delta_{h}=3$ meV. To obtain
analytical results, we choose a constant level width of $\Gamma=0.3$ meV.
We are interested in changes of the character at $\varepsilon_F=0$ on
variation of the Zeeman interaction and hybridization relative to
each other. It must be noted that the $n=0$ LL here plays the most important role.
Figure 1 shows that $C_{Q}$ is zero at $\varepsilon_F=0$. This
occurs because the $n=0$ LL splits only when the hybridization does not vanish.
The $n=0$ LL splits into one electron and one hole level, which reflects a metal to
insulator transition caused by the hybridization (for $\Delta_z=0$). The states are
electron-hole symmetric at this stage. $C_{Q}$ can be tuned from a minimum to a maximum
when the Zeeman energy is increased by changing the external magnetic
field. For $\Delta_z>\Delta_h$ both $n=0$ sublevels are located in the hole region,
which breaks the electron-hole symmetry, see Fig.\ 2, and represents the trivial to
non-trivial topological insulator phase transition.
To observe the splitting, the broadening of the LLs must be less than the
hybridization energy. We note that not only the $n=0$ LL but all LLs are split into
two sublevels.

In Fig.\ 3 we address the SdH oscillations in $C_{Q}$. The following parameters
are used: $\Gamma=0.3$ meV, $T=0$ K, $n_{e}=4\times10^{15}$ m$^{-2}$,
$v=3\times10^{5}$ ms$^{-1}$, and $\Delta_{h}=4$ meV. For low and high magnetic field we
observe a beating pattern and a splitting of the SdH oscillations, respectively.
Both occurs due to interference of the SdH oscillations at two different
frequencies and therefore is a consequence of the splitting of the LLs for
finite Zeeman interaction and hybridization. The beating pattern
vanishes once the Zeeman interaction dominates the hybridization. For
the chosen parameters this occurs at a magnetic field of about 1.5 T.
Above this value we find a well resolved splitting of the SdH oscillations in Fig.\ 3.
A similar beating pattern and splitting of SdH oscillations in the longitudinal
resistivity is seen in Ref.\ \cite{33}, where it occurs because of splitting of the
LLs as a result of a broken inversion symmetry due to substrate effects. The sample
investigated in this reference is thick and hybridization effects are ruled out.

In order to understand the origin of the beating pattern and
splitting of the SdH oscillations quantatively, we employ the Poisson summation
formula
\begin{equation}
\frac{1}{2}F(0)+\underset{k=1}{\overset{\infty}{%
{\displaystyle\sum}}}F(k)=%
{\displaystyle\int\limits_{0}^{\infty}}
F(n)dn+2\overset{\infty}{%
{\displaystyle\sum_{k=1}}}(-1)^{k}%
{\displaystyle\int\limits_{0}^{\infty}}
F(n)\cos[2\pi kn]dn \label{12}%
\end{equation}
to arrive at the following expression for the DOS at zero temperature which
incorporates the contributions of the $n\geq1$ LLs:
\begin{equation}
D(\varepsilon_{F})=D_{0}\left\{  1+2\overset{\infty}{%
{\displaystyle\sum_{\tau_{z},k=1}}
}(-1)^{k}\exp\left[  -\left(  \frac{2\pi k\varepsilon_{F}\sqrt{2}\Gamma
}{\hslash^{2}\omega^{2}}\right)  ^{2}\right]  \cos\left[  \frac{2\pi
k}{\hslash^{2}\omega^{2}}(\varepsilon_{F}^{2}-(\Delta_{z}+\tau_{z}\Delta
_{h})^{2})\right]  \right\}  \label{13}%
\end{equation}
Here, $D_{0}=|\varepsilon_{F}|/\pi\hslash^{2}v^{2}$
is the zero magnetic field DOS. For $\Gamma\gg
\hbar\omega$, it is sufficient to retain only the first order term ($k=1$),
since the higher order terms are highly damped.
From Eq.\ (12) we can confirm that the beating pattern is due to the interference
of waves with two different frequencies. The oscillatory part of the $k=1$
term in the sum for $\tau_{z}=+/-$ is
\begin{equation}
\cos\left[  \frac{2\pi}{\hslash^{2}\omega^{2}}(\varepsilon_{F}^{2}-(\Delta
_{z}+\Delta_{h})^{2})\right]  +\cos\left[  \frac{2\pi}{\hslash^{2}\omega^{2}%
}(\varepsilon_{F}^{2}-(\Delta_{z}-\Delta_{h})^{2})\right]  \label{14}%
\end{equation}
and can be expressed as
\begin{equation}
2\cos\left[  \frac{2\pi}{\hslash^{2}\omega^{2}}(\varepsilon_{F}^{2}-\Delta
_{z}^{2}-\Delta_{h}^{2}{})\right]  \cos\left[  \frac{4\pi}{\hslash^{2}%
\omega^{2}}\Delta_{z}\Delta_{h}\right]. \label{15}%
\end{equation}
This represents a wave of higher frequency whose amplitude oscillates
at a lower frequency, giving rise to the beating pattern, because
$\Delta_{z}\Delta_{h}<<\varepsilon_{F}.$ The amplitude of
the SdH oscillations is modulated by $\cos\left[  \frac{4\pi}{\hslash
^{2}\omega^{2}}\Delta_{z}\Delta_{h}\right]$ and nodes
occur at $\frac{4\Delta_{z}\Delta_{h}}{\hslash^{2}\omega^{2}}%
=\pm0.5,\pm1.5,...$, where the modulating cosine vanishes. We also note
that the amplitude modulation given by the term $\cos\left[  \frac{4\pi
}{\hslash^{2}\omega^{2}}\Delta_{z}\Delta_{h}\right]$ occurs when both
Zeeman interaction and hybridization are present in the system. Moreover,
the threshold magnetic field where beating is seen depends on the strenghts
of the Zeeman interaction and hybridization. For magnetic fields down to
about 1.5 T the beating persists. Above this value it is quenched
to show splitting of the SdH oscillations. As mentioned before this finding
is in agreement with the experimental work in Ref.\ \cite{33}.

\section{Discussion}

Topological systems are typically doped away from the charge neutrality point.
In addition, the bulk usually dominates the surface
contributions. However, thin films of topological insulators make it possible
to access the surface contributions only \cite{16,new1,new2,new3}, typically
when the thickness is below 5 nm. In this
context, we compare our results with the experimental observations
for the quantum magnetocapacitance \cite{xiu2012}. The latter authors study
films that are thick enough that hybridization between the top and bottom surfaces can be
ignored (10 nm Bi$_{2}$Se$_{3}$ samples). To compare their results for a tilted
magnetic field, we next consider this situation and explore
the crossover from external perpendicular to parallel (in-plane) magnetic field on
the quantum capacitance. For a tilt angle $0^\circ\leq\theta\leq90^\circ$ with
the normal of the film, the magnetic field is given by ($B\sin\theta$, 0, $B\cos\theta$).
To describe the effects of the tilting we start from the two-dimensional Hamiltonian
\begin{equation}
H=v\tau_z(\sigma_{x}\boldsymbol{\pi}_{y}-\sigma_{y}\boldsymbol{\pi}_{x})
\end{equation}
where $\mathbf{A}=(0,xB\cos\theta-zB\sin\theta,0)$ and $z$ is the thickness of the film.
The LL energies are obtained as
\begin{equation}
E_{n,\lambda}=\lambda\sqrt{2n}\hslash\omega_{t},
\end{equation}
with $\omega_{t}=v\sqrt{eB\cos\theta/\hslash}$. Note the dependence of the LLs
on the tilt angle $\theta$ as compared to Eq.\ (2).

Following the same procedure as before, we employ the
energy spectrum to compute the quantum capacitance, which is plotted in Fig.\ 4
as a function of the Fermi energy (gate voltage) for different tilt angles. We focus on the
broadening of the LLs and the corresponding effects on the SdH
oscillations. We fix $B=7$ T. The SdH oscillations are suppressed as we increase
the tilt angle, since we are decreasing the perpendicular component of the
magnetic field. For $\theta\to90^\circ$, when the magnetic field is almost
completely aligned with the film, the SdH oscillations are washed out. The broadening
of the LLs generally depends on the magnetic field strength, LL index,
and scattering parameters. This requires a self consistent calculation, which
usually is performed numerically. In order to carry out a tractable analytical
calculation we choose a constant level width of $\Gamma=25$ K. The results in
Fig.\ 4 are consistent with the experimental Fig.\ 4 in Ref.\ \cite{xiu2012}.

An experimental investigation of transport arising from the surface states
in TIs is hindered by the bulk contributions due to naturally occurring
defects and residual carrier doping \cite{rev3,rev4,rev5}. In this regard, the low
mobility of the surface states is a serious problem. However, recent work on TIs
has shown a way to overcome this problem by studying
transport in ultra thin samples, where a 12 times higher surface mobility as compared
to the bulk can be achieved \cite{rev6}. It has been demonstrated that the large
surface to volume ratio of Bi$_2$Se$_3$ thin films effectively suppresses
bulk effects and enables probing of the electronic transport of the surface states \cite{rev7}. 
There has been a rapid improvement in the sample quality coming along with a reduction of
film thicknesses \cite{rev8}. Relevant to the present study, the effect of the Zeeman
exchange field in TIs is significant \cite{new1,rev8} with a magnitude around 50 meV due
to the large effective g-factor. Also, recent experiments have found a
hybridization gap in the ultrathin regime (less than 6 nm) \cite{16,20} as well as quantum
oscillations \cite{new3}. Therefore, it can be expected that the effects discussed before
can be realized in quantum capacitance experiments in ultrathin films.


In conclusion, we have investigated the effects of Zeeman interaction and hybridization
on the magnetocapacitance of a TI thin film in a perpendicular magnetic field. To this aim,
we have obtained an analytic expression for the DOS incorporating both the Zeeman and
hybridization contributions. It turns out that a combination of Zeeman interaction
and hybridization breaks the electron-hole symmetry, resulting in significant effects on the
magnetocapacitance at $\varepsilon_F=0$. In addition, we have shown that in the
presence of both Zeeman interaction and hybridization, the magnetocapacitance
exhibits beating at low and level splitting at high perpendicular magnetic field. This behaviour
is explained by interference of the SdH oscillations at the two frequencies induced by
the splitting of the LLs. Let us comment on the relevance of our findings for experiments.
The cyclotron energy for $B=1$ T is $\hslash\omega=36$ meV. Observation
of the splitting of the LLs and the discussed consequences requires that the
temperature is low and the system clean enough that disorder effects do not wash
out the splitting. Temperature and disorder broadening thus must not reach the Zeeman
or hybridization energies. For a tilted magnetic field with $\theta\to90^\circ$ we find
that the SdH oscillations are washed out.

\begin{acknowledgments}
K.\ Sabeeh would like to acknowledge the support of the Abdus Salam
International Center for Theoretical Physics (ICTP) in Trieste, Italy through
the Associate scheme where a part of this work was completed and the Higher
Education Commission (HEC) of Pakistan for support through project No. 20-1484/R\&D/09.
\end{acknowledgments}

\section*{Author Contributions}
MT performed the calculations. KS contributed to the calculations.
MT and US wrote the manuscript.

\section*{Additional Information}
The authors declare no competing financial interests.

\begin{figure}[h]
\caption{Quantum capacitance as a function of the Fermi energy at $T=0$ K,
$B=3$ T, and Zeeman energy 0 meV. The hybridization energy is 0 meV (dotted
lines) and 4 meV (solid lines), respectively.}
\end{figure}

\begin{figure}[h]
\caption{Quantum capacitance as a function of the Fermi energy at $T=0$ K,
$B=3$ T, Zeeman energy 5 meV, and hybridization energy 3 meV.}
\end{figure}

\begin{figure}[h]
\caption{Quantum capacitance as a function of the magnetic field at $T=0$ K
and hybridization energy 4 meV.}
\end{figure}

\begin{figure}[h]
\caption{Quantum capacitance as a function of the Fermi energy for different
tilt angles of the magnetic field: 0$^\circ$ (dashed lines), 40$^\circ$ (red solid
lines), 70$^\circ$ (blue solid lines), and 80$^\circ$ (black solid lines).
We use $B=7$ T and $\Gamma=25$ K.}
\end{figure}

\end{document}